\documentstyle[12pt,epsfig]{article}
\begin{document}
\begin{titlepage}

\vspace*{-1truecm}

\begin{center}
{\Large \bf  $\epsilon$-EXPANSION IN QUANTUM FIELD  \\
\vskip 0.75 cm
THEORY IN CURVED SPACETIME}
\vskip 0.5 cm
{\large {\bf
Iver H. Brevik${}^{a}$,
Hern\'an Ocampo${}^b$,
Sergei Odintsov${}^{bc}\,$
\footnote{\tt 
Iver.H.Brevik@mtf.ntnu.no,
hocampo@quantum.univalle.edu.co,
odintsov@kakuri2-pc.phys.sci.hiroshima-u.ac.jp,
odintsov@galois.univalle.edu.co,
odintsov@quantum.univalle.edu.co}
}}\\
\vskip 0.75 cm
{\it ${}^a$Department of Applied Mechanics, Norwegian University of 
Science and Technology, N-7034 Trondheim, Norway\\
 ${}^b$Departamento de F\'isica, Universidad del Valle,A.A. 25360 Cali,
Colombia\\
${}^c$Tomsk Pedagogical Universiry, Tomsk, Russia\\
}

\end{center}

\begin{abstract}
We discuss $\epsilon$-expansion in curved space-time for asymptotically
free and asymptotically non-free 
theories. The existence of stable and unstable fixed points is investigated
for $f \phi^4$ theory and $SU(2)$ 
gauge theory. It is shown that $\epsilon$-expansion maybe compatible with
asymptotic freedom on special 
solutions of the RG equations in a special case (supersymmetric theory).
Using $\epsilon$-expansion RG 
technique the effective Lagrangian for covariantly constant gauge SU(2)
field and effective potential for 
gauged NJL-model are found in $4-\epsilon$-dimensional curved space (in
linear curvature approximation). 
The curvature-induced phase transitions from symmetric phase to asymmetric
phase (chromomagnetic 
vacuum and chiral symmetry broken phase, respectively) are discussed for
the above two models.

\end{abstract}
\end{titlepage}
11.10.Hi, 11.30.Qc, 11.15.-q, 04.62.+v

\newpage

\section{Introduction}
The knowledge of infrared structure of quantum field theory is very
important in different applications. For 
example, the confining phase of SU(2) theory  describes the infrared (IR)
structure of the Standard Model. From 
another side, quantum field theory at non-zero temperature(and its infrared
structure) are relevant for 
construction of the inflationary models of the universe.

To study IR properties of the theory, the well-known $\epsilon$-expansion
techniques is often very useful, 
as it happens, for example, in the theory of critical phenomena. For
renormalizable theories the $\epsilon$-
expansion technique gives a standard way to investigate the critical points
of the theory and their stability, 
the critical behavior and the nature of the phase transitions at non-zero
temperature. Of course, $\epsilon$-
expansion works normally well only near $D=4$, i.e. for infinitesimal
$\epsilon$. As usually in such cases, 
one employs standard assumptions that are not strictly correct
mathematically (they are derived for small $\epsilon$) 
but continue to be valid at  $\epsilon$=1.(The main motivation for such
study is, of 
course, the search for finite-temperature phase transition or investigation
of the Casimir effect at non-zero 
temperature). In different areas of physics there exists a variety of
explicit examples (see \cite{2}) where, 
having at our disposal both numerical and experimental data, we conclude
that the comparison of both gives unexpectedly 
good agreement even for $\epsilon \approx 1$.

The present work is devoted to the extension of the $\epsilon$-expansion
technique to curved space-time. Such 
method is clearly required for better understanding of quantum field theory
in the early universe which is hot 
(non-zero temperature) and non-flat one (curvature). And that kind of
theory is necessary for 
construction of the inflationary Universe.

 We start in the next section from the simple discussion of RG equations 
modifications in 4-$\epsilon$-
dimensional curved space, using $f \phi^4$ theory as an example. Working in
the matter sector we describe the 
structure of fixed points of the theory and give the solutions of RG
equations. The extension of this 
discussion for SU(2) gauge models is presented ( in particular, the
structure of fixed points is given 
again). We also show that asymptotic freedom on special solutions of RG
maybe  realized in 4-$\epsilon$-
dimensions only for supersymmetric theory.

In section 3 we consider the calculation of the effective Lagrangian for a
covariantly constant gauge field in 4-
$\epsilon$-dimensions. Actually, that is a particular application of the
$\epsilon$-expansion. The possibility of phase 
transitions induced by the combined effect of curvature and temperature is
discussed.

Section 4 is devoted to the study of chiral symmetry breaking in gauged NJL
model in 4-$\epsilon$-
dimensions, using $\epsilon$-expansion technique and equivalency with
gauge-Higgs-Yukawa model(via 
Bardeen-Lindner-Hill compositeness conditions). The RG improved effective
potential is found and critical 
curvature is defined. Some remarks and outlook are given in the concluding
section.
\section{Renormalization group equations in curved 4-$\epsilon$-dimensional
space-time}
In this section we will be interested in studying the RG behavior of matter
theories in curved 4-$\epsilon$-
dimensional space-time. The typical massless Lagrangian under discussion is
written as follows \cite{1} 
\begin{equation}\label{2.1}
{\cal L} = \mu^{- \epsilon} (a_1 R^2 + a_2 C_{\mu \nu \alpha \beta}^2 + a_3
G) + \frac{1}{2} \xi R \phi^2 
+ {\cal L}_m (\phi, \psi, A_{\mu})
\end{equation}
where G is the Gauss-Bonnet term, $ C_{\mu \nu \alpha \beta}$ is the Weyl
tensor, $\phi$ is a scalar and 
$\mu$ is a mass parameter which is used to make coupling constant to be
dimensionless in 4-$\epsilon$-
dimensions. The symbolic form for the matter Lagrangian which includes
scalars $\phi$, spinors $\psi$ and 
gauge fields $A_{\mu}$ is
\begin{eqnarray}
{\cal L}_m (\phi, \psi, A_{\mu})&=& -\frac{1}{4} G^a_{\mu \nu} G^{a \! \mu
\nu} + \frac{1}{2} g^{\mu 
\nu} (\nabla_{\mu} \phi)^a (\nabla_{\nu} \phi)^a - \frac{1}{4 !} f
\mu^{\epsilon} \phi^4 \nonumber \\
& &+ \bar{\psi}^a i\gamma^{\mu}(x) \nabla_{\mu}^{a b} \psi^b - h
\mu^{\frac{\epsilon}{2}} \bar{\psi} 
\psi \phi \label{2.2}
\end{eqnarray}
Here $\phi^2= \phi^a \phi_a$, $(\nabla_{\nu} \phi)^a, \; (\nabla_{\mu}
\psi)^b$ are covariant derivatives of 
scalar and spinor, respectively. These covariant derivatives include the
curved space covariant derivatives and 
standard gauge coupling term with g changed in the way $g \rightarrow
\mu^{\frac{\epsilon}{2}} g$. Note 
that explicit $\mu$-dependence is introduced in (\ref{2.2}) in order to
keep h, f and g to be dimensionless 
in D=4-$\epsilon$-dimensions. For a moment we consider the gauge group, the
number of scalar and spinor 
multiplets, the features of Yukawa, gauge and scalar interactions to be
arbitrary. However, the theory with 
the Lagrangian (\ref{2.1}) is supposed to be multiplicatively
renormalizable in four-dimensional curved 
spacetime (for details, see \cite{1}).

To study the critical behavior of the system under discussion we will
employ the $\epsilon$-expansion 
technique (for an introduction to $\epsilon$-expansion technique and theory
of critical phenomena, see 
\cite{2,3}). We will be mainly interested in $\epsilon = 1$, i. e. in
finite temperature systems. It is well-
known \cite{4,5} that to study the critical behavior of a system it is
enough to consider only the massless 
subset of the theory. That is why we do not take into account masses in
(\ref{2.1}).

Let us start from the trivial example the $f \phi^4 $-theory in curved
D=4-$\epsilon$-dimensional 
spacetime. We discuss only the matter sector of such a theory. RG flows are
generated by the following RG 
equations:

\parbox{11cm}{\begin{eqnarray*} 
\frac{df}{dt}& =& - \epsilon f +  \frac{3 f^2}{(4 \pi)^2} \label{2.3}\\
\frac{d \xi}{dt} &= &f \frac{( \xi -\frac{1}{6})}{(4 \pi)^2} \label{2.3a} 
\end{eqnarray*}}\parbox{1cm}{\begin{eqnarray}\label{}\end{eqnarray}}

As one can easily see the Eqs. (\ref{2.3}) have the standard form
\cite{4,5}: classical scaling dimension in 4-
$\epsilon$-dimensions plus one-loop $\beta$-function.  Eq. (\ref{2.3a}) for
$\xi$  has again the same 
form as in four dimensions \cite{1}.

The IR behavior of the system defines the critical phenomena. There are two
fixed points of Eqs. \ref{2.3}:

\skip0.5cm
\parbox{11cm}{\begin{tabular}{lll}
1)& Unstable fixed point: & $f^{\ast} = 0, \; \xi^{\ast}$ is arbitrary \\
2)& Stable fixed point:& $f^{\ast} = \frac{(4 \pi)^2 \epsilon}{3}, \;
\xi^{\ast}=\frac{1}{6}$ .
\end{tabular}}\parbox{1cm}{\begin{eqnarray}\label{2.4}\end{eqnarray}}
\vskip0.5cm

The last fixed point for $\xi^{\ast}=\frac{1}{6}$ indicates the phenomenon
of asymptotic conformal 
invariance which was found in \cite{6}. Note that we discuss here only
matter sector. Phase space is more 
extended as the complete fixed point is given by $(f^{\ast}, \;
\xi^{\ast},\;a_1^{\ast},\; 
a_2^{\ast},\;a_3^{\ast})$, ( see also Appendix).

It is quite well-known \cite{2,3,4} that near the stable fixed point of the
$\epsilon$-expansion the system 
experiences second-order phase transitions. At  the same time, first order
phase transitions which are 
typically common in quantum field theory are predicted \cite{5} if the
theory possesses no stable fixed 
point. Moreover, as a rule they are only weakly of first order.

One can also note the explicit solution of the RG equations:

\parbox{11cm}{\begin{eqnarray*}
f(t) & = & \frac{f \epsilon (4 \pi)^2}{3 f - (3 f - \epsilon (4 \pi)^2)
e^{\epsilon t}} \\
\xi (t)- \frac{1}{6} & = &(\xi- \frac{1}{6})(e^{\epsilon t}
\frac{f(t)}{f})^{\frac{1}{3}}
\end{eqnarray*}}\parbox{1cm}{\begin{eqnarray}\label{2.5}\end{eqnarray}}

When $t \rightarrow - \infty $, f(t) $\rightarrow \! f^{\ast}$, $\xi(t)
\rightarrow \! \xi^{\ast}$ corresponding 
to the IR stable fixed point (\ref{2.4}). For a discussion of the
equivalence between the IR behavior of finite-
temperature systems $(\epsilon = 1)$ and three-dimensional theories, see
\cite{7}.

Let us consider now the more interesting example of $SU(2)$ gauge theory
with scalars and spinors. Such a 
model is of importance because the IR behavior of  SM at finite temperature
is described by the confining 
phase of some $SU(2)$ gauge theory with matter. We will be interested in
the theories which are 
asymptotically free \cite{14} on special solutions of RG (\ref{2.1}). The
Lagrangian of such a model is 
given by eq. (3.116) of ref \cite{1} with the evident replacement 
$f \rightarrow f \mu^{\epsilon}$, $h \rightarrow h
\mu^{\frac{\epsilon}{2}}$, $g \rightarrow g 
\mu^{\frac{\epsilon}{2}}$. The theory contains two scalar multiplets and
one spinor multiplet, four scalar 
couplings: $f_1,\! f_2,\! f_3, \! f_4$ and two Yukawa couplings: $h_1,\!
h_2\!$. Using results of \cite{8,6} 
one can explicitly write RGEs for all coupling constants in
4-$\epsilon$-dimensions.  After that, the fixed 
points of the $\epsilon$-expansion can  be  easily found as numerical
solutions of algebraic equations, i.e. zeros 
of the  r.h.s. of RG equations. However, from the very beginning, one finds
that the corresponding system of 
fixed points does not have stable solutions. Only the unstable fixed point
is possible for gauge coupling 
constant $g^2_\ast=0$. Hence only first order phase transitions may be
expected.

We are interested later in the behavior of the effective potential for the
system under discussion. To 
calculate it we need the explicit solution of RG equations. However , even
in exactly four dimensions the 
explicit solution (so-called special solution) of RG equations for the
theory under discussion can be found 
only in two regimes \cite{8}:
\begin{equation} \label{2.6}
h_i^2(t)= k_1^i g^2(t),\;\; f_j(t) =k_2^j g^2(t)
\end{equation}
where numerical constants $k_1^i,\; i=1,2$ and $k_2^j,\; j=1,\cdots,4$ are
different in the above two 
regimes. It is clear that in 
4-$\epsilon$-dimensions the  r.h.s. of the RG equations are modified by
classical scaling of coupling 
constants . The special solution of the type (\ref{2.6}) may survive in
exceptional cases. The theory under 
consideration with Lagrangian (3.116) of \cite{1} belongs to this class. In
one of the regimes of asymptotic 
freedom where the model corresponds to N=2 supersymmetric theory
\begin{equation}\label{2.7}
f_1=f_2=f_3=0,\;\; h_1=h_2=g,\;\; f_4=g^2
\end{equation} 
One can see that this asymptotically free solution is not spoiled by
$\epsilon$-dependent terms of RG equations. In other words, in this regime
4 different RG equations for 
coupling 1s $g^2,\; h_1^2,\; h_2^2,\; f_4$ reduce to the same RG equation:
\begin{equation}\label{2.8}
\frac{d g^2}{d t}=- \epsilon g^2 - \frac{8 g^4}{(4 \pi)^2}
\end{equation}
with the following explicit solution:
\begin{equation}\label{2.9}
g^2(t)= \frac{g ^2}{- \frac{8 g^2}{(4 \pi)^2 \epsilon} + e^{\epsilon t} ( 1
+ \frac{8 g^2}{(4 
\pi)^2\epsilon})}
\end{equation}

For $t \rightarrow \infty ,\; g^2(t)\rightarrow 0$ (asymptotic freedom).

In IR ($t \rightarrow -\infty) ,\; g^2(t)\rightarrow g^2_{\ast} = -\frac{(4
\pi)^2 \epsilon}{8}$. Hence, from 
the analysis of the general solution for coupling constant we also see the
appearance of fixed points in IR 
and UV regions.

We may also ask what happens with the scalar-gravitational coupling
constants $\xi_1, \; \xi_2$. The 
corresponding RG equations are written in the book \cite{1}\newline ( see
Eqs (3.117)) and 
they are the 
same in 4 or in 4-$\epsilon$-dimensions. Taking into
account conditions (\ref{2.7}) we get

\parbox{11cm}{\begin{eqnarray*}
(4 \pi)^2 \frac{d \xi_1 (t)}{d t} & = &  -4 g^2(t) (\xi_1 (t) -
\frac{1}{6}) + 4 g^2(t) (\xi_2(t) - 
\frac{1}{6})\;,\\
(4 \pi)^2 \frac{d\xi_2(t)}{d t}& = & 4 g^2(t) (\xi_1(t) - \frac{1}{6}) -4
g^2(t) (\xi_1(t) -\frac{1}{6})
\end{eqnarray*}}\parbox{1cm}{\begin{eqnarray}\label{2.10}\end{eqnarray}}

Analyzing the r.h.s. of (\ref{2.10}) we see that for the UV stable fixed
point $g_{\ast}^2=0$ the fixed 
points $\xi^{\ast}_1 ,\;\xi^{\ast}_2$ are given by arbitrary values in
agreement with the results of \cite{6}.

For the IR fixed points $g_{\ast}^2$ we get the following: $\xi_1^{\ast} -
\xi_2^{\ast} =0$, otherwise 
$\xi_1^{\ast}$ and $\xi_2^{\ast}$ have arbitrary values. In particular, one
can choose 
$\xi_1^{\ast}=\xi_2^{\ast}=\frac{1}{6}$, to realize the asymptotic
conformal invariance in the IR region.

The explicit solution of Eq. (\ref{2.10}) maybe written as follows:

\parbox{11cm}{\begin{eqnarray*}
\xi_1(t) &=&\frac{1}{2 } (\xi_1 + \xi_2)+\frac{1}{2}(\xi_1 - \xi_2) 
(e^{\epsilon t} 
\frac{g^2(t)}{g^2})^{\frac{1}{2}} \\ 
\xi_2 (t)&=& \frac{1}{2 } (\xi_1 + \xi_2)-\frac{1}{2}(\xi_1 - \xi_2)
(e^{\epsilon t} 
\frac{g^2(t)}{g^2})^{\frac{1}{2}}
\end{eqnarray*}}\parbox{1cm}{\begin{eqnarray}\label{ }\end{eqnarray}}

So far, we discussed the behavior of SU(2) gauge model with matter in
D=4-$\epsilon$-dimensional curved spacetime. In the same manner one
can analyze other gauge models, where unfortunately one can not obtain the
explicit solutions of RGE's. For 
example, let us consider another SU(2) gauge theory with $m$ spinor
triplets, one scalar triplet, one Yukawa 
coupling and one scalar coupling (the Lagrangian is given by Eq.(3.97) from
\cite{1}). The RG equations 
maybe easily written

\parbox{11cm}{\begin{eqnarray*}\label{ }
\frac{d g(t)^2}{d t} & = & -\epsilon g(t)^2 - \frac{(14 - \frac{16}{3} m)
g(t)^4}{(4 \pi)^2}\; ,\\
\frac{d h(t)^2}{d t}& = &-\epsilon h(t)^2 + \frac{16 h(t)^4 - 24 h(t)^2
g(t)^2}{(4 \pi)^2}\; , \\
\frac{d f(t)}{d t}& = & -\epsilon f(t) + \frac{\frac{11}{3}f(t)^2 -24
g(t)^2f(t) +72 g(t)^4}{(4 \pi)^2}\\
 &  & +\frac{16 f(t) h(t)^2 - 96h(t)^4}{(4 \pi)^2}\;, \\
\frac{d \xi (t)}{d t}& = & \frac{1}{(4 \pi)^2}(\xi (t)
-\frac{1}{6})(\frac{5}{3} f(t) + 8 h(t)^2- 12 g(t)^2)
\end{eqnarray*}}\parbox{1cm}{\begin{eqnarray}\label{ }\end{eqnarray}}

This system of RG equations for m=1,2 has only one UV stable point at
$g=0$, h=0, f=0 and $\xi$ arbitrary. 
For $m \geq 3$ we have also an IR stable point at 

$$g^2=-\frac{\epsilon (4 \pi)^2}{14-\frac{16}{3}m}\,,\;\;\;
h^2=-\frac{1}{16} \epsilon (4\pi)^2 
\left(\frac{10-\frac{16}{3}\,m}{14-\frac{16}{3}\,m}\right)\,,$$ 
$$f=-\frac{\epsilon (4\pi)^2}{14-\frac{16}{3}\,m}\left[
\frac{3}{11}(\frac{3}{8}(10+\frac{16}{3}\, 
m)^2-72)\right]^{1/2}\,,\;\;\; \xi=\frac{1}{6}\,.$$

Note finally that by considering interaction of the above models with
renormalizable higher-derivative 
quantum gravity one can estimate the influence of quantum gravity effects
on fixed points of RG in the 
$\epsilon$-expansion \cite{9}.

Another interesting point is related to the possibility of calculating the
RG improved effective potential in 
the above theories in 4-$\epsilon$-dimensions. Let us start from the RG
equations for the effective 
potential (this equation is satisfied due to multiplicative
renormalizability):

\begin{equation}\label{2.11}
(\mu \frac{\partial}{\partial \mu} + \beta_{p_i}\frac{\partial}{\partial
\beta_{p_i}} - \gamma 
\phi\frac{\partial}{\partial \phi}) V(\phi) = 0
\end{equation}

where $p_i$ are all coupling constants, $\beta_{p_i}$ are corresponding
$\beta$-functions and $\gamma$ 
is the anomalous scaling dimension of scalar. Solving (\ref{2.11}) by the
method  of characteristics, using 
tree level potential as boundary condition we obtain the RG improved
effective potential \cite{10} ( for a 
recent discussion of RG improvement in flat and curved space, see \cite{11,
12}, respectively). 
For  
example, for scalar self-interacting theory one gets
\begin{equation}\label{2.12}
V(\phi) = \frac{\mu^{\epsilon} f(t)}{4!}\phi^4 -\frac{1}{2} \xi (t)R \phi^2
\end{equation}
where the running coupling constants are given by Eq. (\ref{2.5}). Taking
$\epsilon =1$ we get leading-log 
behavior of the effective potential in three dimensions. Note that $t=
\frac{1}{2} log 
\frac{\phi^2}{\mu^2}$ where $\mu$ should be identified with the temperature
\cite{13}. Similarly, one can 
find the RG improved effective potential for a SU(2) gauge model or any
other theory.

\section{Effective Lagrangian for a covariantly constant gauge field in
4-$\epsilon$-dimensions}
The $\epsilon$-expansion discussed in the previous section maybe well
applied also to study the effective 
potential for a covariantly constant gauge field in
4-$\epsilon$-dimensional curved space-time. 
In flat D=4 
space such an effective potential (chromomagnetic potential) has been
discussed some time ago in 
\cite{15}. It has been shown that the chromomagnetic potential may have a
nonzero minimum, i.e. there is 
the possibility of chromomagnetic vacuum in an electroweak theory.
Unfortunately, such a state could not be 
the true vacuum of the theory because the effective potential has an
imaginary part \cite{15}. There have 
been various proposals about how this vacuum maybe stabilized.

It is quite possible to expect the presence of large (chromo)-magnetic
fields in early hot universe. There are 
indications that non-zero curvature may lead to some stabilization of the
chromomagnetic vacuum \cite{16}. 
It could be that the combined effect of non-zero temperature and non-zero
curvature may result in the 
vanishing of the imaginary part of the chromomagnetic potential and
stabilization of the corresponding 
vacuum. The $\epsilon$-expansion technique gives the way to calculate the
approximate effective potential 
under such circumstances.

We will consider pure SU(2) gauge theory without matter in a weakly curved
constant curvature space-time 
$R_{\mu \nu}= g_{\mu \nu} R$ where the expansion of the effective potential
over curvature maybe used. 
Taking into account   that the theory is renormalizable, and only the RG
equation for gauge coupling is 
changing due to classical scaling dimension one can repeat the arguments
given in \cite{16} to obtain the 
RG improved effective Lagrangian:
\begin{equation}\label{3.1}
{\cal L}=-\frac{1}{4} \frac{g^2}{g^2(t)} G_{\mu \nu}^a G^{a \; \mu \nu}+
a_1(t) R^2 + a_2 (t) C_{\mu 
\nu \alpha \beta}^2 +a_3 (t) G\; ;
\end{equation}
here $g^2(t)$ has the form similar to (\ref{2.9}) with $\epsilon = 1$. For
SU(2) gauge theory the condition 
of the covariantly constant background is 
\begin{equation}\label{3.2}
\hat{\nabla}^{\mu \; ab} G^b_{\mu \nu}=0
\end{equation}

This equation should be understood as a normal coordinates expansion where
the first term gives 
the flat 
space equation and the remaining ones give curvature corrections. Limiting
ourselves to the case 
of a 
covariantly constant magnetic field we get
\begin{equation}\label{3.3}
\frac{1}{4} G^a_{\mu \nu} G^{a\; \mu \nu} = \frac{1}{2} H^2 + O(R)
\end{equation}

Now, we have to define RG parameter t. For flat space \cite{15}
$t=\frac{1}{2} ln\frac{g H}{\mu^2}$ 
where $\mu$ should be identified with temperature T.

At the same time in curved space-time with vanishing gauge field the
effective mass of the theory is given by 
curvature \cite{1,12}. Hence a possible choice for the RG parameter  t is
\begin{equation}\label{3.4}
t= \frac{1}{2} ln\frac{\frac{R}{4}+ g H}{\mu^2}
\end{equation}

Note that this is very rough estimation as actually we have a mass matrix
which should be properly 
diagonalized. As a result we will get few effective masses and a procedure
of RG improvement with few 
effective masses should be used \cite{17}. However, for a first, rough
estimate the RG parameter t (\ref{3.4}) 
maybe employed. Then we get ($\epsilon =1$) 
\begin{equation}\label{3.5}
{\cal L}= -\frac{1}{2} \frac{g^2}{g^2(t)} H^2 + a_1 (t) R^2 + a_2(t) C_{\mu
\nu \alpha \beta}^2 + a_3(t) 
G
\end{equation}
where
$$g^2(t)= \frac{g^2}{-\frac{11}{12 \pi^2}g^2 + e^{t} ( 1 +
\frac{11}{12\pi^2}g^2)} \;,\; g^2\,< \, 1$$

Due to choice of background $G=\frac{R^2}{6},\;\;\; C_{\mu \nu \alpha
\beta}= 0$ and Eq. (\ref{3.5}) 
maybe rewritten as 
\begin{equation}\label{3.6}
{\cal L}= -\frac{1}{2} \frac{g^2}{g^2(t)} H^2 + (a_1 + \frac{1}{6}a_3 -
\frac{62 \, t}{720(4 \pi)^2}) 
R^2
\end{equation}

The approximate minimum of the effective potential is given by analysis of
the
equation
\begin{equation}\label{3.7}
\frac{ \partial V}{\partial H} = \frac{\partial }{\partial H}(\frac{1}{2}
\frac{g^2}{g^2(t)} H^2)=0
\end{equation}

Numerical analysis of the effective potential maybe done.

For example, in D=4 ($\epsilon =0$) one gets \cite{16}
\begin{equation}\label{3.8}
g H_{min} = \mu^2 \exp{\frac{-24 \pi^2}{11 g^2}} - \frac{R}{4}
\end{equation}
Curvature slightly modifies the (unstable) chromomagnetic vacuum of
\cite{16}, and may even act in the 
direction of its stabilization \cite{15}.

At non-zero temperature ($\epsilon = 1$) and curvature one obtains:

a) At zero curvature
\begin{equation}\label{3.9}
g H_{min} = T^2 (\frac{11 g^2}{15 \pi^2})^{\frac{1}{2}}
\end{equation}

So we find the possibility of temperature-induced phase transitions.
However, the non-zero vacuum state is unstable \cite{15} what can not be
seen in our approximation. The 
qualitative form of the potential is given in Fig. 1.

b) At non-zero curvature (R $\ll$ g H) we get
\begin{equation}\label{3.3}
(\frac{g H_{min} + \frac{R}{4}}{ T^2})^{\frac{1}{2}} = \frac{\frac{11
g^2}{15 \pi^2} + \sqrt{(\frac{11 
g^2}{15 \pi^2})^2 + \frac{R}{5 \pi^2}}}{2}
\end{equation}

Hence we find here the possibility of phase transitions induced by the
combined effect of curvature and 
temperature. The schematic behavior of the effective potential is again the
same as above. The critical line 
on the phase plane R-T is given by Eq. (\ref{3.3}) at $H_{min}=0$.

The present picture indicates that for some values of curvature and
temperature the imaginary part is 
canceled by the combined effect of non-zero curvature and temperature
(stabilization of chromomagnetic 
vacuum). However to check this conjecture one would have to consider the
above problem not in 
$\epsilon$-expansion but exactly (making explicit calculation of effective
Lagrangian at non-zero T and 
some fixed gravitational background).

\section{Gauged NJL model in 4-$\epsilon$-dimensional curved space-time}
In the present section we will consider gauge-Higgs-Yukawa theory of the
type (\ref{2.2}). We work in 
modified $1/N_c$-expansion in 4-$\epsilon$-dimensional curved space-time
and we use the equivalence of 
the above model with gauged NJL model (what was shown in \cite{22}). That
gives  the way to study gauged 
NJL-model  in 4-$\epsilon$-dimensions.

We start from the $SU(N_c)$ gauge theory with scalars and spinors in 
4-$\epsilon$-dimensional curved 
space-time:
\begin{eqnarray}
{\cal L}_m + \frac{1}{2 }\xi R G^2 &=&-\frac{1}{4} G^a_{\mu \nu} G^{a\;\mu
\nu} + \frac{1}{2} 
g^{\mu \nu} \partial_{\mu} \sigma \partial_{\nu}\sigma -\frac{1}{2} m^2
\sigma^2-\frac{1}{2} \lambda \mu^{\epsilon} 
\sigma^4\nonumber \\ 
& &+\frac{1}{2} \xi R \sigma^2
 + \sum_{i=1}^{N_f} \bar{\psi_i} i \gamma^{\mu} \nabla_{\mu} \psi_i
-\sum_{i=1}^{n_f}y 
\mu^{\frac{\epsilon}{2}} \sigma \bar{\psi_i} \psi_i\label{4.1}
\end{eqnarray}
where $\sigma$ is a single scalar, $\lambda$ is scalar coupling, $N_f$
fermions belong to the representation R of 
$SU(N_c)$ and we denote the Yukawa constant by y here as in \cite{22}.

The modified $\frac{1}{N_c}$-approximation of \cite{22} consists in
choosing  small gauge coupling 
(first non-trivial order in $g^2$), $N_f \sim N_c$ but $n_f \ll N_f$, and
dropping of scalar loop 
contributions (leading order of $\frac{1}{N_c}$). Within such an approach
and taking also the classical 
scaling dimensions into account we get
\parbox{11cm}{\begin{eqnarray*}\label{4.2}
\frac{dg(t)}{dt}& = &-\frac{\epsilon}{2} g(t) - \frac{b g^3(t)}{(4 \pi)^2
}\;\; ,\\
\frac{d y(t)}{dt}&=& y(t) \left[\frac{a y^2(t)}{(4 \pi)^2} - \frac{c
g^2(t)}{( 4 \pi)^2} - 
\frac{\epsilon}{2}\right] \;\;,\\
\frac{d \lambda(t)}{dt}& = &u y^2(t)\left[\frac{\lambda(t)}{(4 \pi)^2}
-\frac{y^2}{(4 \pi)^2}\right] -
\epsilon \lambda \;\;,\\
\frac{d\xi(t)}{dt}& =& \frac{1}{(4 \pi)^2}2 a y^2(t)(\xi(t)-\frac{1}{6})
\;\;,
\end{eqnarray*}}\parbox{1cm}{\begin{eqnarray}\label{4.2}\end{eqnarray}}

where \cite{22}: $b=\frac{11N_c-4 T(R)N_f}{3}$, $c=6C_2(R)$,
$a=\frac{u}{4}=2n_f N_c$. Here 
$t=ln\frac{\mu}{\mu_0}$, and for the fundamental representation
$T(R)=\frac{1}{2}$, 
$C_2(R)=\frac{N_C^2-1}{2 N_c}$. Only the case $c>b$ will be considered.
Note that the extension of above 
model to the case of external gravitational and magnetic fields has been
done in \cite{23,24} respectively. 
Solving the RG system (\ref{4.2}) one can find that this system of RG 
equations only leads to an UV stationary point at g=0, y=0, $\lambda=0$,
and $\xi$ arbitrary.

One can easily obtain the solution of RG equations in terms of RG
invariants:

\parbox{11cm}{\begin{eqnarray*}
\frac{g^2(t)}{g^2}& \equiv & \eta (t)\equiv \frac{\alpha (t)}{\alpha} =
\frac{\epsilon}{-\frac{2bg^2}{(4 \pi)^2} + (\epsilon + \frac{2b g^2}{(4
\pi)^2})e^{\epsilon t}}\; ,\\
y^2(t) & =& \frac{c-b}{a} g^2(t) [1 + h_0 (e^{\epsilon t} \eta
(t))^{1-\frac{c}{b}} ]^{-1} \\
\lambda(t) & = &\frac{2a}{2c -b} \frac{y^4(t)}{g^2(t)} [1 + k_0
(e^{\epsilon t} \eta(t))^{1-
\frac{2c}{b}}] \; , \\
\xi (t)- \frac{1}{6}& = & (\xi - \frac{1}{6}) \frac{y^2(t)}{y^2}
(e^{(1-c/b)\epsilon t} 
\eta^{-c/b}(t))
\end{eqnarray*}}\parbox{1cm}{\begin{eqnarray}\label{4.3}\end{eqnarray}}

where the RG invariants $h_0\;, k_0$ are defined as:

\parbox{11cm}{\begin{eqnarray*}
h(t)&=&-(e^{\epsilon t} \eta (t))^{-1+ \frac{c}{b}} [1 -
\frac{c-b}{a}\frac{g^2(t)}{y^2(t)}]\;,\\
k(t)&=&-(e^{\epsilon t} \eta (t))^{-1+\frac{2c}{b}}\left[1-\frac{2c-b}{2a}
\frac{\lambda 
(t)}{y^2(t)}\frac{g^2(t)}{y^2(t)}\right]\;.
\end{eqnarray*}}\parbox{1cm}{\begin{eqnarray}\label{4.4}\end{eqnarray}}

(They dont depend on t so we may add subscript zero).

Below, we only consider the case of fixed gauge coupling ($b\rightarrow
+0$):
\begin{equation}\label{4.5}
(e^{\epsilon t} \eta(t))^{\frac{-c}{b}} \longrightarrow
 \exp \left\{\frac{\alpha}{\alpha_c}\left( \frac{1-e^{- \epsilon
t}}{\epsilon}\right)\right\}
\end{equation}
where $\alpha^{-1}_c=\frac{3c_2(R)}{\pi}$, $\alpha=\frac{g^2}{4 \pi}$ and

\parbox{11cm}{\begin{eqnarray*}
y^2(t) &=&\frac{c-b}{a} g^2(t) \left[1+h_0 \exp
\left\{\frac{\alpha}{\alpha_c}\left(\frac{1- e^{-\epsilon 
t}}{\epsilon}\right)\right\}\right]^{-1} \\
\lambda(t)&=& \frac{2a}{(4 \pi)^2} \frac{\alpha_c}{\alpha}y^4(t)\left[
1+k_0 \exp 
\left\{\frac{2\alpha}{\alpha_c}\left(\frac{1 - e^{-\epsilon
t}}{\epsilon}\right)\right\}\right]
\end{eqnarray*}}\parbox{1cm}{\begin{eqnarray}\label{4.6}\end{eqnarray}}

The analysis of  the behavior of the above coupling constants at $\epsilon
=0$ has been done in \cite{22} 
where phase structure has been studied with all details.

Now we turn to the $SU(N_c)$ gauged NJL model with four-fermion coupling
constant $\tilde{G}$ in curved space-
time:
\begin{equation}\label{4.7}
{\cal L}= -\frac{1}{4} G_{\mu\,\nu}^{a\; 2} + \sum_{i=1}^{N_f} \bar{\psi}_i
i \gamma^{\mu}(x) 
\nabla_{\mu} \psi_i + \tilde{G}\sum_{i=1}^{n_f}(\bar{\psi_i}\psi_i)^2\; ,
\end{equation}
$\tilde{G}$ let denote four-fermion coupling.

The easiest way to study such a model is to introduce an auxiliary field
$\sigma$, in order to identify the NJL-
model with the Higgs-Yukawa model. As has been shown by
Bardeen-Hill-Lindner \cite{25} one can put a 
set of boundary conditions (compositness condition) for the effective
couplings of the gauge-Higgs-Yukawa 
model at $t_{\Lambda}=\ln \frac{\Lambda}{\mu_0}$ (where $\Lambda$ is UV
cutoff of the gauged NJL 
model) in order to prove  the equivalence of the  gauged NJL model with the
gauge-Higgs-Yukawa model 
\cite{22}. Then one can study gauged NJL model using the RG method as for
usual renormalizable theories 
but the RG method is more easy than the complicated Schwinger-Dyson
equation \cite{27}.

In our discussion using compositeness conditions for coupling constant we
may actually define RG 
invariants. Then, omitting the details which are very similar to ones given
in \cite{25,22}(without 
$\epsilon$-dependence), one may use Eqs. (\ref{4.3}), (\ref{4.6}) and find
the running Yukawa and scalar 
couplings in the gauged NJL model (g(t) is not changing). We write them
below for $b \rightarrow +0$:

\parbox{11cm}{\begin{eqnarray*}
y^2(t)& = & y^2_{\Lambda}(t) = \frac{(4 \pi)^2}{2a}
\frac{\alpha}{\alpha_c}\left[ 1- \left(\frac{\exp 
\left(\frac{1- (\mu_0/\mu)^{\epsilon}}{\epsilon}\right)}{\exp
\left(\frac{1-
(\mu_0/\Lambda)^{\epsilon}}{\epsilon}\right)}\right)^{\frac{\alpha}{\alpha_c
}}\right]^{-1} \\
\frac{\lambda(t)}{y^4(t)} &=
&\frac{\lambda_{\Lambda}(t)}{y^4_{\Lambda}(t)}=\frac{2a}{(4 \pi)^2} 
\frac{\alpha_c}{\alpha}\left[ 1
- \left(\frac{\exp \left(\frac{1-
(\mu_0/\mu)^{\epsilon}}{\epsilon}\right)}{\exp\left(\frac{1
-(\mu_0/\Lambda)^{\epsilon}}{\epsilon}\right)}\right)^{\frac{2
\alpha}{\alpha_c}}\right]
\end{eqnarray*}}\parbox{1cm}{\begin{eqnarray}\label{4.8}\end{eqnarray}}

where $t<t_{\Lambda}$. In the limit $\epsilon \rightarrow 0$ Eqs.
(\ref{4.8}) coincide with the  
corresponding Eqs of  \cite{22}.

It is interesting to note that in the limit $\Lambda \rightarrow \infty$ we
get from (\ref{4.8}):

\parbox{11cm}{\begin{eqnarray*}
y^2_{\Lambda}(t) &\longrightarrow & \frac{(4 \pi)^2}{2a}
\frac{\alpha}{\alpha_c} \left[1 - \left\{exp 
\left(\frac{-\left(\mu_0/\mu\right)^{\epsilon})}{\epsilon}\right)\right\}^{\
alpha/\alpha_c} \right]^{-1}\\
\frac{\lambda_{\Lambda}(t)}{y^4_{\Lambda}}& \longrightarrow& \frac{2a}{(4
\pi)^2} 
\frac{\alpha_c}{\alpha} \left[ 1 - \left\{ exp \left(-
\frac{(\mu_0/\mu)^{\epsilon}}{\epsilon}\right)\right\}^{2
\alpha/\alpha_c}\right]
\end{eqnarray*}}\parbox{1cm}{\begin{eqnarray}\label{4.9}\end{eqnarray}}

In addition to compositeness conditions for scalar and Yukawa couplings, in
order to prove the equivalence 
between gauge-Higgs-Yukawa model (\ref{4.1}) and
gauged NJL model (\ref{4.7}) one should also have the compositeness
condition for mass and for scalar-
gravitational coupling constant.

The analysis of the compositeness condition for mass is completely similar
to the one given in \cite{22}. 
So we present only the result in the limit $b \rightarrow +0$:

\begin{equation} \label{4.10}
m^2(t)= \frac{2a}{(4 \pi)^2} y^2_{\Lambda}(t)\left( exp \left\{
\frac{(\mu_0/\Lambda)^{\epsilon}-
(\mu_0/\mu)^{\epsilon}}{\epsilon}\right\}\right)^{\alpha/\alpha_c}
\Lambda^2 
\left[\frac{1}{g_4(\Lambda)}- \frac{1}{w}\right]
\end{equation}

where $g_4(\Lambda)$ is a dimensionless constant defined by
$\tilde{G}\equiv \left(\frac{(4 
\pi)^2}{a}\right) \frac{g_4(\Lambda)}{\Lambda^2}$ and $w=1- \frac{\alpha}{2
\alpha_c}$, 
$y^2_{\Lambda}(t)$ is given by Eq (\ref{4.8}).

The compositness condition for $\xi(t)$ is the same as in non-gauged NJL
model \cite{26} (see also 
discussion in \cite{23} for gauged NJL model):

\begin{equation}\label{4.11}
\xi(t)= \frac{1}{6} 
\end{equation}

Thus using the equivalence with the gauge-Higgs-Yukawa theory we get the
description of the gauged NJL 
model via running coupling constants.

Having this description of the gauged NJL model via renormalizable
SU($N_c$) gauge theory one can apply 
RG improvement technique to calculate the effective potential (taking
account of leading logarithms from 
the whole set of diagrams). Following the method of  \cite{22} (and its
extension to curved space-time 
\cite{23}) one can solve the RG equation for effective potential in
following form:

\begin{equation}\label{36}
V(g,\,y,\,\lambda,\,m^2,\,\xi,\,\sigma,\,\mu)=
V(\bar{g}(t),\dots,\bar{\sigma}(t),\mu e^t)
\end{equation}

Here the effective coupling constants $\bar{g}(t),\dots,\bar{\sigma}(t)$
are defined by RG equation 
(\ref{4.2}) (at scale $\mu e^t$) and  RG eqs. for $\bar{m}^2(t),\,
\bar{\sigma}(t)$ are given in 
\cite{22}.As the boundary condition it is convenient to use the one-loop
effective potential \cite{23}.

For gauged NJL model we have to substitute to RG improved potential the
effective couplings which fulfill 
the compositeness conditions (see discussion above). Evaluating the running
parameters at t determined by

\begin{equation}\label{37}
\bar{{\cal M}_F}(t)\equiv \bar{y}(t) \bar{\sigma}(t)= e^t \mu
\end{equation}

one should insert (\ref{37}) into the effective potential (\ref{36}). From
the RG invariant (at b 
$\longrightarrow \; +0$) and (\ref{37})

\begin{equation}\label{38}
\exp \{ \frac{\alpha}{\alpha_c} \frac{1-e^{-\epsilon t}}{\epsilon} \}
\bar{{\cal M}}_F(t) 
\end{equation}

one defines the connection between t and ${\cal M}_F$:

\begin{equation}\label{39}
\exp \{\frac{\alpha}{\alpha_c} \frac{1-e^{-\epsilon t}}{\epsilon}\} e^{2
t}= \frac{{\cal 
M}_F^2(\mu)}{\mu^2}
\end{equation}

Using the fact that $\bar{m}^2(t) \bar{\sigma}^2(t)$ is also RG invariant,
applying (\ref{38}), (\ref{39}) and 
(\ref{4.10}), (\ref{4.11}) we can find the quadratic part of the RG
improved effective potential (for related 
discussion, see \cite{22,23,24})

\parbox{11cm}{\begin{eqnarray*}
V_2& =& \frac{1}{2} \frac{2 a}{(4 \pi)^2} y^2_{\Lambda}(\mu)\left( 
exp\{\frac{(\frac{\mu_0}{\Lambda})^{\epsilon}-
(\frac{\mu_0}{\mu})^{\epsilon}}{\epsilon}\}\right)^{\alpha/\alpha_c}\Lambda^
2 
\left[\frac{1}{g_4(\Lambda)}- \frac{1}{w}\right] \sigma^2(\mu)\\ 
& &+\frac{R {\cal M}_F^2(\mu)}{12\,\bar{y}^2(t)}+ \frac{a R (e^t \mu)^2}{12
(4 \pi)^2}
\end{eqnarray*}} \parbox{1cm}{\begin{equation}\label{40}\end{equation}}

Here, $\bar{y}^2_{\lambda}(t)= y^2(e^t \mu)$ is the rescaling of $y^2(t)$.
It is obtained from the   (\ref{4.9}) 
where one takes $e^t \mu$ instead of $\mu$. The factor $e^t$ is fixed by
the relation (\ref{39}).

 In a similar way one can find the four-scalar part of the effective
potential which is not important for us. 
That is the quadratic part which defines the symmetry breaking:

\begin{equation}\label{41}
V_2 < 0
\end{equation}

For $V_2 < 0$ the symmetry is broken. Even in flat space there is the
possibility for chiral symmetry 
breaking. In curved space-time, the condition $V_2=0$ defines the critical 
curvature which gives the critical 
line between the symmetric phase and the chiral symmetry broken phase.
Hence, we again showed the principal 
possibility of curvature-induced phase transition \cite{1} in gauged NJL
model in curved 4-$\epsilon$-
dimensional space-time.

\section{Discussion}
In the present work we have developed the $\epsilon$-expansion technique
for quantum field theory in 
curved spacetime. Unlike the case of flat space, the phase space of the
present theory is extended. One has additional RG 
equation for the scalar-gravitational coupling constant $\xi$ and for
vacuum coupling constants and the 
corresponding fixed points.

We discussed the possibility for survival of asymptotic freedom of all
coupling constants as special 
solution of the RG equations in 4-$\epsilon$-dimensions. This phenomenon
may happen only with 
supersymmetric theory (in flat space), unlike to the case of four
dimensions. As applications of 
the $\epsilon$-expansion technique we found the RG improved effective
Lagrangian for covariantly constant gauge 
field in SU(2) theory and the RG improved effective potential for gauged
NJL-model in curved space with non-
zero temperature. The phase structure of both models was discussed. It is
very interesting to note that such 
calculation of the RG improved effective action in curved space with
non-zero temperature may find 
cosmological applications (for example, in the calculation of the wave
function of the universe).

Another interesting line of research is related with the application of the
$\epsilon$-expansion in the study 
of Casimir effect at non-zero temperature in solid state physics. Here, one
again has to find some RG 
improved effective action. (We give few remarks on Casimir effect in media
in Appendix B).

Finally, let us note that there exists a long-standing proposal of Weinberg
\cite{18} to formulate consistent, 
asymptotically-safe quantum gravity in $2+\epsilon$-dimensions. The
subsequent analytic continuation to 
$\epsilon=1$ or $\epsilon=2$ is then necessary. (Note that it is quite
probable that the program of Weinberg 
maybe consistently realized in dilatonic gravity \cite{19,21}, at least up
to the point before analytic 
continuation). From this point of view, further study of the
$\epsilon$-expansion in 4-$\epsilon$-dimensional 
curved spacetime (and subsequently, in 4-$\epsilon$-dimensional quantum
gravity) may provide us with new 
ideas, as some synthesis of 2+$\epsilon$-dimensional approach with
4-$\epsilon$-dimensional approach 
could lead to coinciding results for $\epsilon=1$. At least, such analysis
may enrich both approaches.

{\large {\bf Acknowledgments}}. The work by SDO has been supported in part
by COLCIENCIAS 
(Colombia).
\section{Appendix A: Running vacuum couplings}
In the present Appendix we describe the behavior of vacuum effective
couplings in 4-$\epsilon$-
dimensions.

For vacuum (or external fields) effective couplings in $f \phi^4$-theory we
get (compare with \cite{28})

\parbox{11cm}{\begin{eqnarray*}
\frac{d \, a_1}{d\,t} & =& \epsilon a_1 + \frac{1}{2 (4 \pi)^2} (\xi -
\frac{1}{6})^2 \\
\frac{d \, a_2}{d\,t} & =& \epsilon a_2 + \frac{1}{120 (4 \pi)^2} \\
\frac{d \, a_3}{d\,t} & =& \epsilon a_3 - \frac{1}{360 (4 \pi)^2} 
\end{eqnarray*}}\parbox{1cm}{\begin{eqnarray}\label{A.1}\end{eqnarray}}

The general solution of Eqs. (\ref{A.1}) is :

\parbox{11cm}{\begin{eqnarray*}
a_1(t) &=& \left[ a_1 + \frac{(\xi -\frac{1}{6})^2}{6f} \left(ln
(4\pi)^2\epsilon +\epsilon t +ln 
\frac{f(t)}{(4 \pi)^2\epsilon f} \right)\right] e^{\epsilon t}\\
a_2(t)&=& -\frac{1}{120(4\pi)^2 \epsilon}+ \left[ a_2 +\frac{1}{120(4
\pi)^2\epsilon}\right] e^{\epsilon 
t} \\
a_3(t)&=& \frac{1}{360(4\pi)^2 \epsilon}+ \left[ a_3 -\frac{1}{360(4
\pi)^2\epsilon}\right] e^{\epsilon 
t} \end{eqnarray*}} \parbox{1cm}{\begin{eqnarray}\label{A.2}\end{eqnarray}}

At the IR fixed point (when $t\longrightarrow\, -\infty$) we get 

\parbox{2.5cm}{$$a_1(t)\, \longrightarrow\,0\;,$$}  \parbox{4cm}{$$\,a_2(t)
\longrightarrow\,-
\frac{1}{120 (4 \pi)^2\epsilon}\;,$$}  \parbox{4cm}{$$\, a_3(t) 
\longrightarrow\,\frac{1}{360(4\pi)^2\epsilon}\;,$$}\parbox{1.5cm}{
\begin{equation}\label{A.3}\end{equation}}

Similarly, one can find surface coupling constants \cite{28,31} in
4-$\epsilon$-dimensions. It is also easy to 
generalize the above RG equations for more complicated theories, the
qualitative structure of solutions 
(\ref{A.2}) will be the same.

One can apply the RG equations (\ref{A.2}) to find RG improved effective
action $\Gamma$ on quasi-De 
Sitter background $S_3\times S_1$: $R_{\mu\, \nu}= \Lambda \,
g_{\mu\,\nu}$.

To this end one can write the RG improved effective action on $S_4$ 
\cite{28} (at zero background scalar)

\begin{equation}\label{A.4}
\Gamma = 24 \pi^2 \{16 a_1(t) + \frac{8}{3} a_3(t)\}
\end{equation}

Now, in order to translate this expression to $\Gamma$ on $S_3\times S_1$
one should use $a_1(t)$ and 
$a_2(t)$ given by Eqs. (\ref{A.2}) (with $\epsilon =1$ and $t=\frac{1}{2}
ln \frac{4\Lambda}{\mu^2})$. 
Note that $\mu$ should be identified with the temperature \cite{13}. Of
course, such approximation is 
rather qualitative as we consider $\epsilon=1$.

Note also that the above result (\ref{A.4}) provides a good example of the
calculation of gravitational Casimir 
effect (for a good introduction to the theory of Casimir effect, see
\cite{29,30})
\section{Appendix B: On the calculation of the Casimir energy in realistic
media}
We close our work by making some remarks on the Casimir energy under
realistic physical conditions 
emphasizing the electromagnetic case. Note that this Appendix is somehow
outside of the general line of 
discussion.

First of all, one must bear in mind that any boundary surface is made up of
molecules of a real medium. The 
medium possesses in general both dispersive and absorptive properties. The
boundary conditions 
that are 
conventionally adopted in scalar field theory, namely the Dirichlet or the
Neumann conditions, can never be 
regarded as anything else than purely idealized conditions that lose their
validity at extreme frequencies. The 
discussion is most conveniently carried out in the electromagnetic case,
where the simplest boundary 
conditions are those of perfectly conducting surfaces. As long as one
confines oneself to a "normal"  
frequency interval, it is in most cases quite legitimate to take the
surfaces to be perfectly conducting. 
However, once  the frequency w becomes much higher than the absorption
frequency $w_p$ (we consider 
only one absorption frequency, for simplicity), then the medium behaves
like a plasma. For a normal metal 
such as copper, $w_p=5\times10^{16}\, s^{-1}$. For $w>>w_p $ the material
gradually becomes more 
"soft",  and finally ceases to have any influence on the photons. In this
way, nature itself provides us with a 
dispersive, soft UV cutoff. Analogous considerations apply to the scalar
case: Dirichlet or 
Neumann 
conditions ceases to exist at sufficiently high frequencies. The material,
and thus also boundaries, gradually 
become soft and fade out of recognition for the field quanta.

As typical example in electrodynamics, let us consider the calculation of
Casimir energy E for a spherical 
vacuum cavity of radius a in a dielectric medium ( this example is closely
related to the phenomenon of 
sonoluminescence). It is natural in this context to calculate E via the
Casimir surface for F 
which acts at the 
boundary surface r=a. This force, in turn, is equal to the difference
between the radial diagonal components 
of  the Maxwell stress tensor at $r = a  \pm $. Now, there has been some
discussion in the recent literature 
about how to regularize the expression for F. There is actually a need for
two kinds of regularization here: 
first, we have to subtract from the Green function the pure volume
contributions. That is at r=a- we subtract 
the Green function corresponding to the inner region (vacuum) filling all
space, whereas at r=a+ 
we subtract 
the Green function corresponding to the outer region (dielectric) filling
all space. This implies that the 
calculated expression for F goes to zero when $r\longrightarrow\, \infty$,
and makes it accordingly for us to 
derive the Casimir energy via the formula

\begin{equation}
F= - \frac{1}{4 \pi a^2}\frac{\partial\, E}{\partial \,a}\; .
\end{equation}

This is the method of calculation given, for instance, by Milton and Ng
\cite{32} -cf. also \cite{33}.

The need of a second kind of regularization stems from the fact that the
expression for E diverges when 
summed over angular momenta l. It has turned out a very convenient way of
obtaining the nondispersive  part 
if the energy (or force) is to make use of the {\it Riemann zeta function}.
This approach was followed in 
\cite{32,33}.

 So what is the role of dimensional regularization in the context?
Dimensional regularization, generally 
speaking, is a calculations device in which singularities in calculated
expressions can be isolated. It 
therefore becomes clear that this kind of regularization may serve as a
substitute for the second step above, 
i. e. , a substitute for the Riemann zeta function method \cite{34}.
Although it would be of interest to 
calculate the Casimir energy for the cavity by means of dimensional
regularization, we are not aware that 
such a calculation has been done.

So far, we have ignored gravitation. If a gravitational field is present,
the analysis goes through in the same 
manner as above, provided that the gravitational radius is much larger than
the intermolecular spacing in the 
material. Then all phenomenological parameters, such as permitivity or
permeability, can still be used. By 
contrast, if the field becomes so strong that gravitational radius is of
the same order as the intermolecular 
spacing, then we have leave the simplest macroscopic picture of the
material and resort to ma much more 
complicated many-body picture. It would be very interesting to include the
temperature to above 
consideration (for example, via the $\epsilon$-expansion technique).

\end{document}